\documentstyle[multicol,aps,prl,graphicx,amssymb]{revtex}
\draft

\begin{document}

\newcommand{\beq}{\begin{eqnarray}}
\newcommand{\eeq}{\end{eqnarray}}
\renewcommand{\thefootnote}{\fnsymbol{footnote}}

\title{Ensemble averages and nonextensivity at the edge of chaos of one-dimensional maps}

\author{Garin F.J. Ananos$^{1,2}$ and %\thanks{fedorja@cbpf.br}
Constantino Tsallis$^{1}$ %\thanks{tsallis@cbpf.br}
}
\address{
$^1$Centro Brasileiro de Pesquisas F\'\i sicas, \\
Rua Dr.  Xavier Sigaud 150, %\\
22290-180 Rio de Janeiro, RJ, Brazil \\
$^2$Departamento de F\'{\i}sica, Universidad Nacional de Trujillo \\
Av. Juan Pablo II, s/n, Trujillo, Per\'u \\
}
\date{\today}
\maketitle
\begin{abstract}
Ensemble averages of the sensitivity to initial conditions $\xi(t)$ and the entropy production per unit time of a {\it new} family of one-dimensional dissipative maps, $x_{t+1}=1-ae^{-1/|x_t|^z}(z>0)$, and of the known logistic-like maps, $x_{t+1}=1-a|x_t|^z(z>1)$, are numerically studied, both for {\it strong} (Lyapunov exponent $\lambda_1>0$) and {\it weak} (chaos threshold, i.e., $\lambda_1=0$) chaotic cases. In all cases we verify that (i) both $\langle \ln_q \xi \rangle \;[\ln_q x \equiv (x^{1-q}-1)/(1-q); \;\ln_1 x=\ln x]$ and $\langle S_q \rangle \;[S_q \equiv (1-\sum_i p_i^q)/(q-1);\; S_1=-\sum_i p_i \ln p_i]$ {\it linearly} increase with time for (and only for) a special value of $q$,  $q_{sen}^{av}$, and (ii) the {\it slope} of  $\langle \ln_q \xi \rangle$ and that of   $\langle S_q \rangle$ {\it coincide}, thus interestingly extending the well known Pesin theorem. For strong chaos, $q_{sen}^{av}=1$, whereas at the edge of chaos, $q_{sen}^{av}(z)<1$. 

\noindent
PACS number: 05.20.-y, 05.45.-a, 05.45.Ac, 05.45.Df %\\
% 05.20.-y : Classical Statistical Mechanics 
% 05.45.-a : Nonlinear dynamics and nonlinear systems
% 05.45.Ac : Low-dimensional chaos
% 05.45.Df : Fractals
% 05.45.+b : Theory and models of chaotic systems
\end{abstract}

\begin{multicols}{2}

%\newpage

Low-dimensional non-linear maps play an important role in the development of the theory of chaos in physics and mathematics. They exhibit various routes to chaos and their related metric universality classes \cite{cvitanivic}. In particular, one-dimensional maps are paradigmatic models to study the emergence of complexity in dynamical systems. In this field, much work has been done to find properties of chaos that could enable a classification of deterministic systems. In particular, dynamical indicators like sensitivity to initial conditions, Lyapunov exponents, Kolmogorov-Sinai (KS) entropy \cite{kolsinai,hilborn}, topological entropy and others have been developed. Nevertheless, it is known today that there are ubiquitous natural and artificial dynamical systems for which these standard indicators give but a poor description of the complexity of their time evolution. This is typically the case in the frontier between standard chaos and regular orbits. Such complex deterministic behaviour has been termed {\it weak chaos}, to distinguish it from {\it strong} or {\it full} chaos. The latter correspond to positive metric entropy or, equivalently, positive Lyapunov exponent. The former occurs for zero values of these indicators. This phenomenon has led to the generalization of the involved concepts \cite{TPZ}.

A new and fruitful approach that characterizes these weakly chaotic dynamical systems is related to the nonextensive generalization \cite{CT1988,CTBJP} of the Boltzmann-Gibbs (BG) statistical mechanics. This generalization has since more than one decade raised much interest in physical situations that do not satisfy the customary BG thermal equilibrium conditions, e.g., quick mixing in phase space and ergodicity. In such cases, anomalous dynamical properties are the rule: see  \cite{turbulentbeck,Latora-Rapisarda-CT,Giansanti}, among many others.

In the present paper, we study numerically the {\it sensitivity to initial conditions} and the {\it entropy production per unit time} (a concept completely analogous to the Kolmogorov-Sinai entropy rate) of the standard $z$-logistic maps as well as of a new family of one-dimensional dissipative maps, from now on referred as to the {\it exponential} maps, here introduced in order to have one more paradigmatic universality class on which basic properties can be tested. In all the calculations we present here, we make {\it ensemble averages} by sampling randomly the entire phase space. This procedure  will be shown to substantially modify the behavior at the edge of chaos, whereas nothing very drastic is changed for strong chaos. These results lead us to the numerical confirmation, for both families of maps, of an appropriate generalization of the well known Pesin identity (i.e., the coincidence of the positive Lyapunov exponent and KS entropy rate). This generalization was conjectured in 1997 \cite{TPZ}, recently proved for special trajectories (more precisely as upper bounds of all trajectories; see \cite{fulvio} for the logistic maps), and is here verified, for the first time, for ensemble averages. 

Let us first recall the $z$-logistic family of maps 
\begin{equation}
\label{map}
x_{t+1} = 1 - a |x_t|^z \;\;\; (z>1; a \in [0,2]; |x_t| \le 1).
\end{equation}
The critical value $a_{c}(z)$ ({\it chaos threshold} or {\it edge of chaos}) monotonically increases from $1$ to $2$ when $z$ increases from 1 to infinity. The $z=2$ map is, as well known, isomorphic to $y_{t+1}\propto  y_{t}(1-y_{t})$. The $z$-logistic maps constitute important universality classes of unimodal maps. The larger $z$ is, the flatter is the  $x=0$ maximum. To characterize a degree of flatness that  $z \to \infty$ cannot attain, we introduce the following new family of maps, inspired in Cauchy's exponential function (infinitely differentiable at $x=0$ and nevertheless nonanalytic):
\beq
\label{map1}
x_{t+1}=1-ae^{-1/|x_t|^z}(z > 0;  a \in [0, a^*(z)];  |x_{t}| \le 1) \,,
\eeq 
where $a^*(z)$ depends slowly from $z$ (e.g., $a^*(0.5) \simeq 5.43$). We only address here $z \ge z_c \simeq 0.4$, $z_c$ being the value above which the attractors are topologically isomorphic (see Fig. 1) to those of the logistic map. 
In fact, one observes a whole cascade of period doublings $2^{k-1} \rightarrow 2^k$ at parameter values $a_{k}$ that accumulate at $a_{c}(z)$ (e.g., $a_{c}(0.5)=3.32169594...$) above which chaos exists.

\begin{figure}[htb]
\begin{center}
\includegraphics[width=0.45\textwidth,keepaspectratio,clip=]{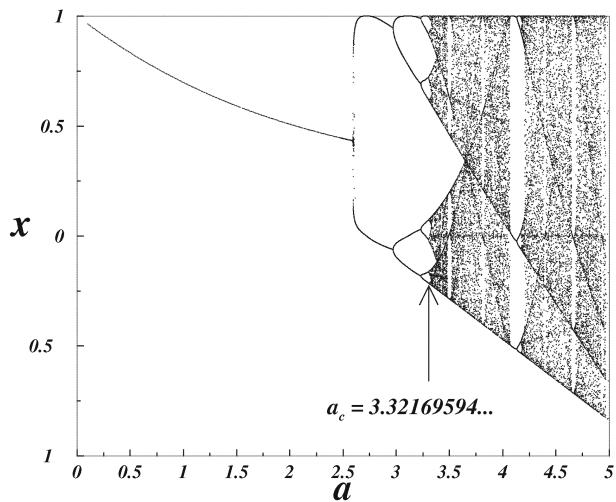}
\end{center}
\caption{\small
$a$-dependence of the dynamical attractor of the $z=0.5$ exponential map. 
}
\label{Fig_1}
\end{figure}

\begin{center}
{\bf Table-} Logistic (top) and exponential (bottom) maps. 
%{\bf Table $1$}
\begin{tabular}{|c|c|c|c|c|c|}
\hline 
&$z$ & $a_c$ & $q^{av}_{sen}$ & $\lambda^{av}_{q^{av}_{sen}}$ & $K^{av}_{q^{av}_{sen}}$ \\ \hline
&$\;1.10\;$ & $\;1.12498...\;$ & $\;0.43 \pm 0.01\;$ & $\;0.12 \pm 0.01\;$ &$\;0.12 \pm 0.02\;$ \\ 
\hline
&$\;1.25\;$ & $\;1.20951...\;$ & $\;0.42 \pm 0.01\;$ & $\;0.18 \pm 0.01\;$ &$\;0.18 \pm 0.02\;$ \\ 
\hline
&$\;1.50\;$ & $\;1.29550...\;$ & $\;0.40 \pm 0.01\;$ & $\;0.22 \pm 0.01\;$ &$\;0.22 \pm 0.02\;$ \\ 
\hline
&$\;1.75\;$ & $\;1.35506...\;$ & $\;0.37 \pm 0.01\;$ & $\;0.26 \pm 0.01\;$ &$\;0.26 \pm 0.02\;$ \\ 
\hline
&$\;2.00\;$ & $\;1.40115...\;$ & $\;0.36 \pm 0.01\;$ & $\;0.27 \pm 0.01\;$ &$\;0.27 \pm 0.02\;$ \\ 
\hline
&$\;2.50\;$ & $\;1.47055...\;$ & $\;0.34 \pm 0.01\;$ & $\;0.28 \pm 0.01\;$ &$\;0.28 \pm 0.02\;$ \\ 
\hline
&$\;3.00\;$ & $\;1.52187...\;$ & $\;0.32 \pm 0.01\;$ & $\;0.29 \pm 0.02\;$ &$\;0.29 \pm 0.03\;$ \\ 
\hline
&$\;5.00\;$ & $\;1.64553...\;$ & $\;0.28 \pm 0.01\;$ & $\;0.30 \pm 0.02\;$ &$\;0.30 \pm 0.03\;$ \\ 
\hline
\hline
&$\;0.40\;$ & $\;3.05996...\;$ & $\;0.40 \pm 0.01\;$ & $\;0.25 \pm 0.01\;$ &$\;0.25 \pm 0.02\;$ \\ 
\hline
&$\;0.50\;$ & $\;3.32169...\;$ & $\;0.35 \pm 0.01\;$ & $\;0.27 \pm 0.01\;$ &$\;0.27 \pm 0.02\;$ \\ 
\hline
&$\;0.75\;$ & $\;3.68229...\;$ & $\;0.30 \pm 0.01\;$ & $\;0.29 \pm 0.02\;$ &$\;0.29 \pm 0.02\;$ \\ 
\hline
&$\;1.00\;$ & $\;3.90705...\;$ & $\;0.25 \pm 0.02\;$ & $\;0.32 \pm 0.02\;$ &$\;0.32 \pm 0.02\;$ \\ 
\hline
&$\;1.25\;$ & $\;4.07088...\;$ & $\;0.20 \pm 0.02\;$ & $\;0.39 \pm 0.02\;$ &$\;0.39 \pm 0.03\;$ \\ 
\hline
&$\;1.50\;$ & $\;4.19820...\;$ & $\;0.15 \pm 0.02\;$ & $\;0.43 \pm 0.02\;$ &$\;0.43 \pm 0.03\;$ \\ 
\hline
\end{tabular}
\end{center}

The $z$-logistic maps, and many others, have already been deeply studied in the regions of strong chaos. The sensitivity to the initial conditions $\xi(t) \equiv \lim_{\Delta x(0) \to 0} \frac{\Delta x(t)}{\Delta x(0)}$, where $\Delta x(0)$ is the discrepancy of initial conditions at time $t=0$ and $\Delta x(t)$ its time dependence, satisfies the differential equation $d\xi/dt=\lambda_1\xi$, where $\lambda_1$ is the Lyapunov exponent, hence $\xi=e^{\lambda_1 t}$. Consequently, if $\lambda_1<0$ ($\lambda_1>0$), the system is said to be {\it strongly insensitive} ({\it sensitive}) to the initial conditions (and intermediate rounding). If, however, $\lambda_1=0$ (as it happens at the edge of chaos), then $\xi(t)$ is expected to satisfy  a more general equation, namely $d\xi/dt=\lambda_q \xi^q$ ($\lambda_{q}$ being a $q$-generalized Lyapunov coefficient), hence \cite{TPZ}
\beq
\label{sensitivity}
\xi = [1+(1-q)\lambda_q \, t]^{\frac{1}{1-q}} 
\equiv e_q^{\lambda_q t} \,.
\eeq
This function, {\it $q$-exponential}, recovers the exponential function  (a power law) for $q=1$ ($q \ne 1$.) If $q < 1$  and $\lambda_{q} > 0$ ($q>1$ and $\lambda_{q} < 0)$, the system is said to be {\it weakly sensitive} ({\it insensitive}) to the initial conditions.

Consistently with these results, whenever strong chaos is present, the appropriate entropy is known to be the BG one, $S_{BG}=-\sum_{i=1}^W p_i \ln p_i$. But, at the edge of chaos, $S_{BG}$ becomes poorly informative, and we must instead use (\cite{TPZ,CT1988,Costa-Lyra-Plastino-CT,Lyra-CT,Tirnakli-Tsallis-Lyra} and references therein)   
\begin{equation}
\label{qentropy}
S_q = \frac{1- \sum_{i=1}^W p_i^q}{q-1}\;\;\; (S_1=S_{BG}) .
\end{equation}
This nonextensive entropy enables in fact the generalization \cite{CT1988} of BG statistical mechanics itself. For each universality class of edge of chaos, one special value of the entropic index $q$ is to be used, which we note $q_{sen}$, where {\it sen} stands for {\it sensitivity}. 
Up to now, four different methods have been developed which provide the special value $q_{sen}$. The first method is based on the sensitivity to initial conditions \cite{TPZ,Costa-Lyra-Plastino-CT}, the second one is related to the entropy production per unit time \cite{Latora-Baranger-Rapisarda-CT,Tirnakli-Ananos-CT}, the third one is based on the geometrical description of the multifractal attractor \cite{Lyra-CT}, and the last one concerns relaxation properties \cite{ernesto}. We address here the first two methods. For the $z$-logistic map, $q_{sen}(z)$ is known for typical values of $z$ \cite{Costa-Lyra-Plastino-CT,Lyra-CT}: See filled circles in Fig. 3 (e.g., $q_{sen}(z)=0.2445...$; also \cite{fulvio} $\lambda_{0.2445}= \ln \alpha / \ln 2=1.3236...$, where $\alpha$ is the Feigenbaum constant). But, for the present exponential family, this is still unpublished and out of the scope of the present work. Our main goal here is to study what happens when we do {\it averages} (noted $\langle...\rangle$) over the entire phase space (i.e., $-1 \le x \le 1$), instead of using the special region around $x=0$ (location of the maximum of the map). 

For the sensitivity we proceed as follows. We consider, at an initial location of $X$, two points very close (say at a distance $\Delta x(0)=10^{-12}$) and then numerically calculate $\xi(t)$ from its definition. We do this operation many times, starting from values of $x$ randomly chosen in the allowed interval. Finally, we average all the values of $\ln_q \xi$, where the inverse function of the $q$-exponential function is defined as follows:
\begin{eqnarray}
\ln_{q}x\equiv\frac{x^{1-q} -1}{1-q}  \;\;(\ln_1 x=\ln x)\,.
\label{qlog}
\end{eqnarray}
Various values of $q$ are tested until we obtain a {\it linear} time dependence of $\langle \ln_q \xi \rangle(t)$. That is the special value of $q$ we are looking for, and we denote it $q_{sen}^{av}$, where {\it av} stands for {\it average}; we denote $\lambda^{av}_{q_{sen}^{av}}$ the linear coefficient. See Fig. 2. For example, for the edge of chaos of the $z=2$ logistic map, we obtain   $q_{sen}^{av} \simeq 0.36$, in remarkable contrast with  $q_{sen} \simeq 0.2445$. We shall come back onto this point later. No such discrepancy is obtained for strong chaos, where we verify in all cases that  $q_{sen}^{av}= q_{sen}=1$. The same method is applied to the exponential family of maps: See Fig. 2 (e.g.,  at the edge of chaos $a_{c}(0.5)=3.32169594$, we obtain $q_{sen}^{av}(0.5) \simeq 0.35$). 

\end{multicols}

\begin{figure}[htb]
\begin{center}
\includegraphics[width=0.65\columnwidth,keepaspectratio,clip=]{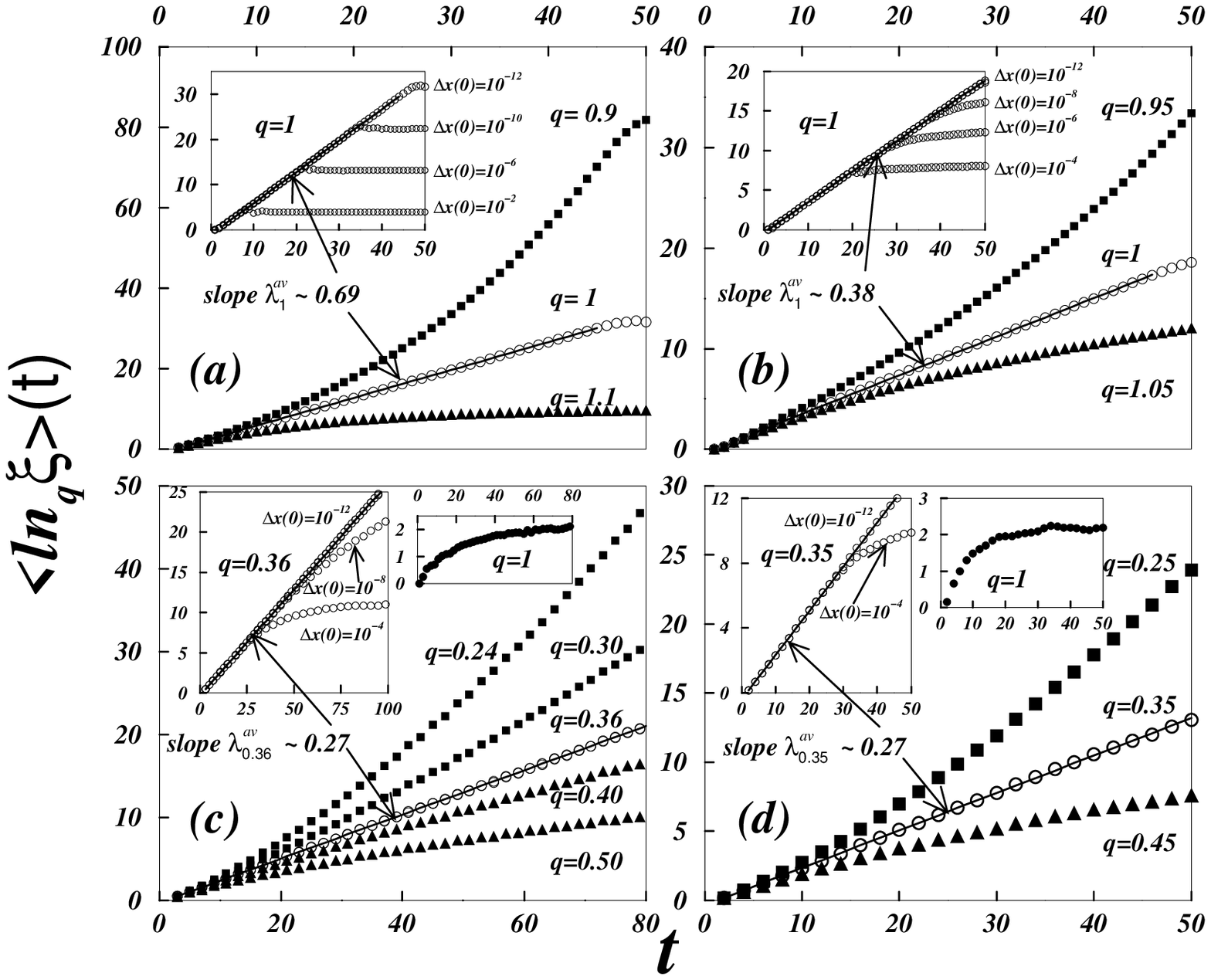}
\includegraphics[width=0.65\columnwidth,keepaspectratio,clip=]{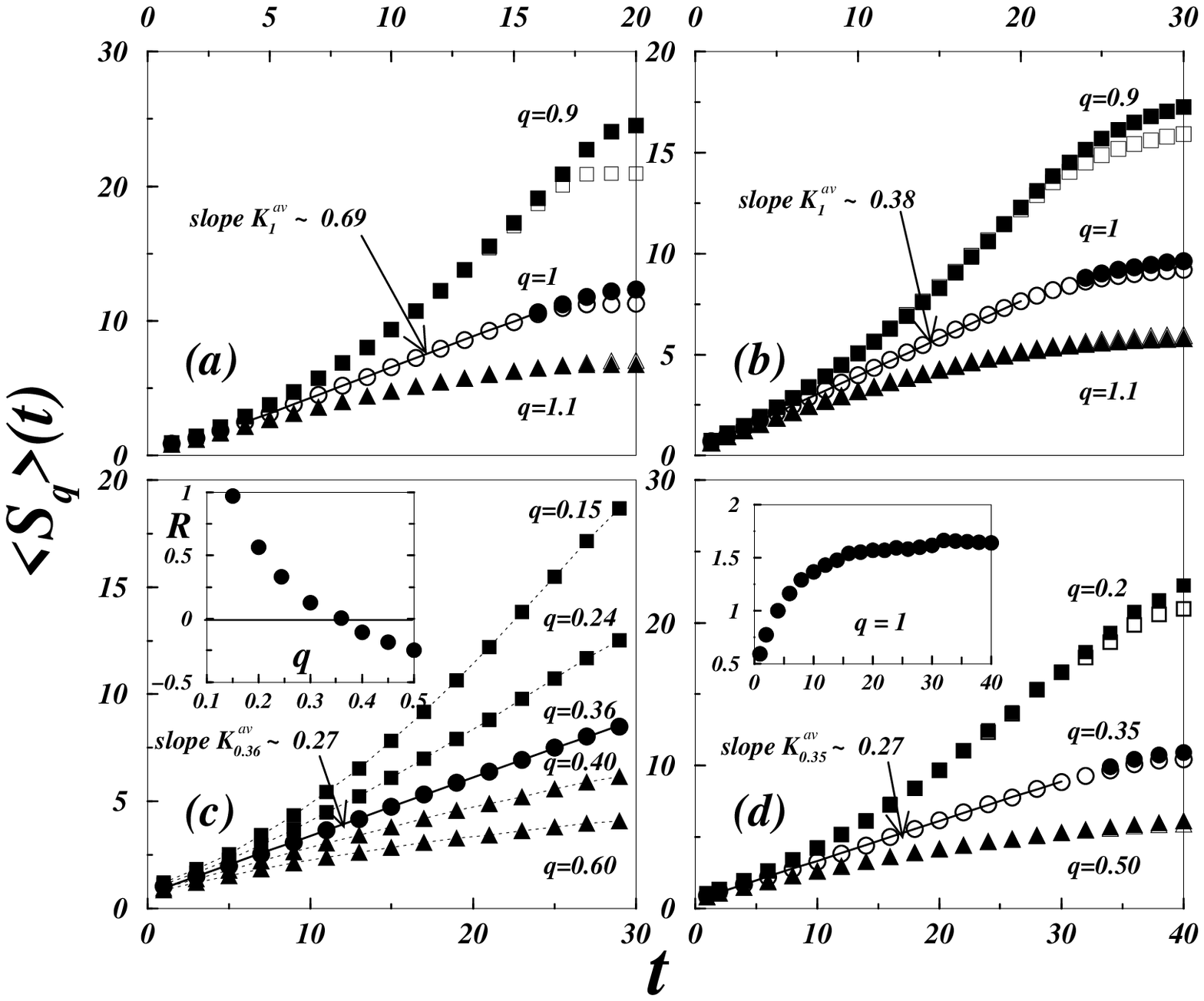}
\end{center}
\caption{\small Time dependence of $\langle \ln_q \xi \rangle$ and $\langle S_q \rangle$:  $z=2$ logistic map for strong [(a) $a=2$]  and weak [(c)  $a=1.401155189$] chaos, and $z=0.5$ exponential map for strong [(b) $a=4$] and weak [(d) $a= 3.32169594 $] chaos. 
{\bf Sensitivity function $\langle \ln_q \xi  \rangle(t)$:} 
averages over $10^5$ ($10^7$) runs for $(a)$ and $(b)$ ($(c)$ and $(d)$); we use $\Delta x(0)=10^{-12}$ as the initial discrepancy unless otherwise indicated; in the insets, we show the {\it linear} tendency of the sensitivity function for $q^{av}_{sen}$ with various values of $\Delta x(0)$; at the edge of chaos ($(c)$ and $(d)$) we exhibit the $q=1$ curve {\it nonlinearity}.  {\bf Entropy $\langle S_q \rangle(t)$:} (a,b) $3000$ runs with $N=10W$ with  $W=10^5$  and $W=3.10^5$ (empty  and filled symbols respectively); 50000 runs with $N=10W$ with  $W=10^5$  (for (c)) and $W=5.10^4$  and $W=10^5$ (for (d)). (c) inset: determination of $q_{sen}^{av}$ (see text).  (d) inset: we exhibit the $q=1$ curve {\it nonlinearity}.
}
\label{Fig_2}
\end{figure}

\begin{multicols}{2}

\begin{figure}[htb]
\begin{center}
\includegraphics[width=0.45\textwidth,keepaspectratio,clip=]{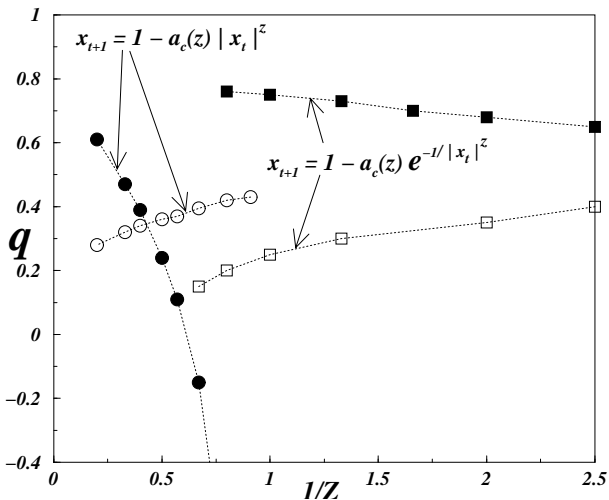}
\end{center}
\caption{\small 
$z$-dependence of $q^{av}_{sen}$ (empty circles and squares: present work) and $q_{sen}$ (filled circles: from \protect \cite{Costa-Lyra-Plastino-CT,Lyra-CT}; filled squares: from \protect \cite{unpublished}). Dotted lines are guides to the eye.
}
\label{Fig_10}
\end{figure}

For the entropy production we proceed as introduced in \cite{latbar} for conservative systems and then used in \cite{Latora-Baranger-Rapisarda-CT} for the logistic map. We divide the phase space in $W$ equal intervals, and put initially (randomly or not) $N$ points in one of them. We then accompany the spread of points within phase space, and calculate $S_q(t)$. This enables the calculation of the entropy production per unit time
\begin{equation}
K_q \equiv \lim_{t\rightarrow\infty} \lim_{W\rightarrow\infty}
\lim_{N\rightarrow\infty} \frac{S_q(t)}{t}
\end{equation}
We then repeat many times starting from randomly chosen intervals and finally average the entropy $S_q$ itself. We test various $q$'s until  we obtain a {\it linear} time dependence of $\langle S_q \rangle(t)$; we note $K^{av}_{q_{sen}^{av}}$ the linear coefficient. It turns out that in all cases we obtained the {\it same} value $q_{sen}^{av}$ provided  by the method of the sensitivity  (see Fig. 3). And in all cases we obtained  $K^{av}_{q_{sen}^{av}}=\lambda^{av}_{q_{sen}^{av}}$, which remarkably extends the standard Pesin theorem \cite{pesin}. 

This kind of averaging, on one hand mimics Gibbs' approach of thermostatistical ensembles, on the other hand 
is a natural way to minimize the considerable fluctuations which appear at the edge of chaos $a=a_{c}(z)$. At this threshold, and in contrast with the case where strong chaos exists, considerably large fluctuations appear in say $S_{q}(t)$. In order to get rid of these fluctuations, we average $\langle S_{q} \rangle (t)$ over a large enough number of initial windows instead of only using the ``best windows" as done in \cite{Latora-Baranger-Rapisarda-CT,Tirnakli-Ananos-CT}. Here, we typically used $W/2$ windows chosen randomly. The fluctuations that so remain are considerably smaller. 

In the chaotic regime for the $z=2$ logistic map we obtained, for $a=2$, the well known result $\lambda_1=K_1=\ln 2 \simeq 0.69$. In the chaotic regime ($a=4$) for the $z=0.5$ exponential map, we obtained $\lambda_1 \simeq K_1 \simeq 0.38$. 

In order to make the entropy result more precise (see the inset (c) of Fig. 2 (bottom) for the standard logistic map as illustration) we fitted the curves $\langle S_{q} \rangle(t) $ in the time interval $[t_{1},t_{2}]$ with the polynomial $A+Bt+Ct^2$. We define the nonlinearity measure $R \equiv C(t_{1}+t_{2})/B$ ($R=0$ for a perfect straight line). For the logistic map ($z=2$), we choose $(t_{1},t_{2})=(1,30)$ for all $q$'s, so that the factor $(t_{1}+t_{2})$ is just a normalizing constant. This inset shows that $R$ vanishes for $q=0.36 \pm 0.01$. These results are not sensitive to changes in $t_{2} \le 30$.  We used the same procedure for the exponential maps. 

Let us summarize our main results: (i) We introduced a {\it new universality class} for one-dimensional unimodal dissipative maps, Eq. (2), which corresponds to a proper characterization of extremely flat maps ($z \to \infty$ of Eq. (1)); (ii) For this family of maps as well as for the $z$-logistic ones, we performed {\it Gibbs-ensemble-like averages} on both
the sensitivity and of the entropy production, and determined (exclusively from dynamics) the entropic index $q_{sen}^{av}(z)$, {\it one and the same from both procedures} (as it is known to occur for $q_{sen}(z)$); (iii) For {\it strong chaos}, we verify $q_{sen}^{av}(z)=q_{sen}(z)=1$ and $K^{av}_{q_{sen}}(z)=\lambda^{av}_{q_{sen}}(z)=K_{q_{sen}}(z)=\lambda_{q_{sen}}(z)$; (iv) At the {\it edge of chaos}, we verify that $q_{sen}^{av}(z)$ {\it decreases for increasing flatness} (i.e., increasing $z$, in contrast with $q_{sen}(z)$), and $K^{av}_{q_{sen}}(z)=\lambda^{av}_{q_{sen}}(z) < K_{q_{sen}}(z)=\lambda_{q_{sen}}(z)$; the fact that generically $q_{sen}^{av}(z) \ne q_{sen}(z)$ is a direct consequence of the fact that different initial conditions yield strongly fluctuating trajectories which follow along the path of virtually any other trajectory {\it but at shifted times} \cite{fulvio} (no such phenomenon exists for strong chaos). 
All these results are expected to contribute to the correct interpretation of experimental results in dynamical complex systems, in particular the ubiquitous dissipative ones. 

We acknowledge interesting remarks from F. Baldovin, E.P. Borges, A. Robledo and U. Tirnakli, as well as partial financial support from PRONEX/MCT, CNPq and Faperj (Brazilian agencies).

\end{multicols}

\end{document}